\documentclass[preprint,aps,preprintnumbers,longbibliography,longbibliography]{revtex4-1}

\usepackage[latin9]{inputenc}
\usepackage{amsmath}
\usepackage{amssymb}
\usepackage{graphicx}
\usepackage{esint}
\usepackage[scr=boondoxo, scrscaled=1.05]{mathalfa}
\allowdisplaybreaks

\makeatletter

\pdfpageheight\paperheight
\pdfpagewidth\paperwidth

\pdfoutput=1

\usepackage{epsfig}
\usepackage{graphics}

\usepackage{mciteplus}
\mciteErrorOnUnknownfalse

\makeatother

\begin{document}

\title{Symmetries of the  Primordial Sky}

\author{Craig  Hogan}
\affiliation{University of Chicago}

\author{Essay written for the Gravity Research Foundation 2022 Awards for Essays on Gravitation}
\address{Corresponding author: Craig Hogan, University of Chicago, 5640 S. Ellis Ave., Chicago, IL 60637, craighogan@uchicago.edu }
\date{\today}
\begin{abstract}
Quantum field theory, which is generally used to describe the origin of  large-scale gravitational perturbations during cosmic inflation, has been shown to omit an important physical effect in curved space-time, the nonlocal  entanglement among quantized modes from their gravitational effect on  causal structure.
 It is argued here that in a different model of  quantum gravity that  coherently preserves nonlocal directional and causal relationships,
 primordial perturbations originate instead from coherent quantum distortions of emergent inflationary horizons;  and moreover, that causal constraints  account for approximate  symmetries of cosmic microwave background correlations measured at large angular separations, which are highly anomalous in the standard picture. Thus,  symmetries  already apparent in  the  large-angle CMB  pattern may be unique  signatures of the emergence of locality and causal structure from quantum gravity.
\end{abstract}
\maketitle


The world works according to two all-encompassing theories:  quantum theory, which describes the behavior of matter and energy, and general relativity, which describes the behavior of space and time, including gravity.  Each one appears  to be,  in its own realm,  a practically flawless description of nature.

Although they describe the same physical world, the foundations of these theories seem to be fundamentally incompatible.  The exact nature  of their conflict has been expressed in many different  ways, depending on different formulations of quantum mechanics.  Locality in space and time stops making any kind of sense on a very small scale;   a quantum system smaller than the Planck length,
$l_P=\sqrt{\hbar G/c^3}
=1.6\times 10^{-35}$  meters, 
 is more compact than a black hole of the same mass.    But the  fundamental incompatibility is much broader and deeper than that, and is not confined to  small systems\cite{HollandsWald2004,Stamp_2015}.
 The indeterminate consequences of a  quantum event spread everywhere at the speed of light, and its  nonlocal gravitational  effect  distorts  space-time, including its causal structure, in all directions and on all scales.
  In practice, context-dependent
approximations must be used to reconcile quantum coherence with geometrical locality and causality.



For most purposes, it is possible to set aside this  knotty  problem, since the active quantum effects of  gravity are very small in almost all physical systems we can actually measure. A notable exception is  cosmology: during cosmic inflation, the  gravitational effects of quantum fluctuations  left  permanent distortions in the large scale structure of space and time, which ultimately led to   the large scale structure  we observe today.  In this sense, quantum gravity is the origin of all cosmic structure, and its effects are precisely measured.  Maps of the cosmic microwave background (CMB) on  large angular scales preserve a largely intact image of the pattern of gravitational potential  on  our cosmic horizon generated by quantum processes during inflation.

The standard quantum theory of how structural perturbations form is a dynamical quantum field theory that includes gravity, which I will call simply the ``QFT model''.  In this picture, a   
  classical (unquantized) cosmological  space-time expands exponentially, due to the gravitational effect of a bespoke scalar ``inflaton'' field that has a large expected vacuum energy density. On this homogeneous classical background, which rapidly  magnifies the physical size of all structures,  a quantum field theory is introduced to compute the quantum fluctuations of the inflaton, and its gravitational effects on  geometrical perturbations that persist, after inflation is over, to create cosmic structure. QFT achieves  localization of quantum states  by borrowing it from the classical background:  mode amplitudes and phases are quantized, each one like a harmonic oscillator, but their mapping onto events in space-time is classical.
  
  The    QFT  model has been the basic framework adopted in the theory of cosmic inflation since it was invented. 
    It is clear  why this framework has been used  for about 40 years: the powerful tools of effective field theory  enable extensive and detailed calculations for many different models of  matter fields during inflation, and lead to a cosmological model that can account for a great deal of precise data, especially on angular scales smaller than a few degrees.

Of course,  the QFT model has not actually solved the basic problems of quantum gravity, it just hides them.
An elegant essay by Hollands and Wald \cite{HollandsWald2004} explains its subtle flaw: the quantized distortion of causal structure  from gravity on all scales is not correctly taken into account in the QFT model, since the fields are defined using  a fixed classical background with a determinate causal structure. 
In the real system, differences in causal structure created by modes with wavenumber $k\rightarrow 0$, including events at  distance $R\rightarrow \pm \infty$,
entangle them with quantized modes on smaller scales. 
 The renormalization of a field theory requires  subtracting infinities to describe the  physical modes of the vacuum; but with gravity included, as pointed out by Hollands and Wald\cite{HollandsWald2004}, 
{\it ``An individual mode will have no way of knowing whether its own subtraction is correct unless it ``knows'' how the subtractions are being done for all the other modes.'' } 
The  QFT mode decomposition models  locality by  acausally (and unphysically) constraining the system at infinity independently in all directions. 
During inflation, acausal phase correlations between null waves arriving from antipodal  directions
at $R=\pm \infty$ are built into the {\it a priori} classical definitions of the quantized stationary comoving modes. This assumption matters for  locality, because the actual physical position of a particle along a line depends on phase information  arriving from opposite directions.

Thus, we should not be surprised that the QFT framework, by ignoring important gravitational quantum entanglement with causal structure on large scales, leads to well-known  gravitational paradoxes outside of inflation, including infrared inconsistencies\cite{CohenKaplanNelson1999}, the apparent loss of information in black-hole evaporation, and a wildly incorrect estimate\cite{Weinberg1988} for the value of the cosmological constant. Its great success in microscopic experiments, especially high energy particle collisions, apparently does not extrapolate to arbitrarily extended quantum systems  where gravity is important. 
I  argue here that the  QFT approximation also leads to incorrect predictions when applied to cosmic inflation, in particular,  for large-angle correlations on horizons.



\begin{figure*}
  \centering
  \includegraphics[width=0.45\textwidth]{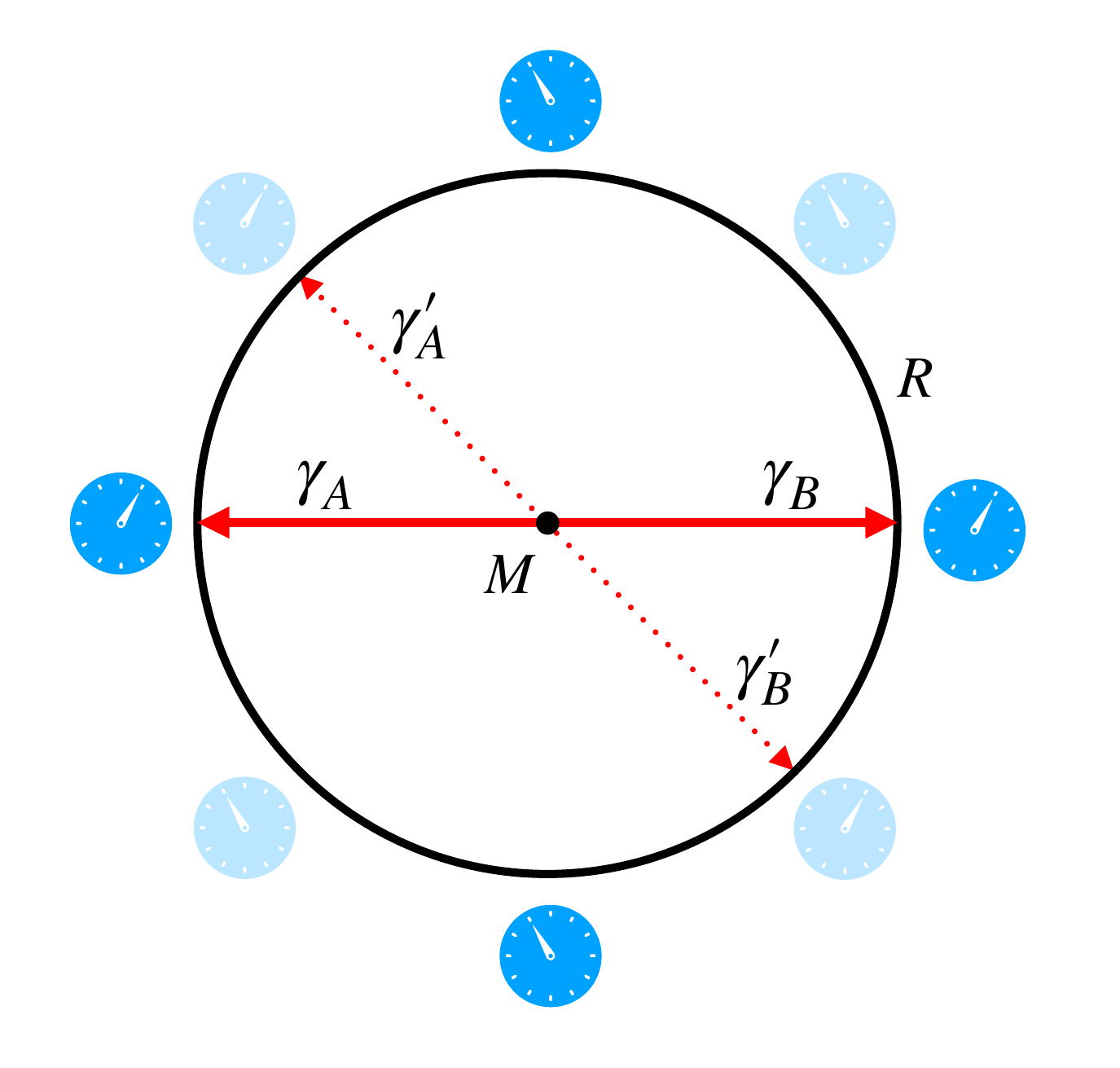}
  \caption{A particle of mass $M$ decays into two photons $\gamma_A, \gamma_B$ that travel in opposite directions.  Their gravitational shock wave distorts  time as measured by observing clocks from the origin, by an amount $\delta \tau\sim GM/c^3$, independent of distance $R$, in a coherent pattern (Eq. \ref{distortion}) aligned with the particle axis, as shown.   An  $S$- wave decay  isotropically superposes states with different particle axes; one other example,  $\gamma_A', \gamma_B'$, is shown.   The gravity of this state places the space-time into a Schr\"odinger-cat-like macroscopic superposition of  different states,  with causal structures that differ coherently on large scales by $\delta \tau\sim GM/c^3$.   An observer at the center measures the same coherent large-scale angular pattern for a causal diamond  of any size.\label{decay}}
\end{figure*}

As an example of the gravitational effect of coherent causal entanglement on an angular pattern, 
 consider the classical distortion of  causal structure from the gravity of  an  idealized EPR-type  system\cite{Mackewicz2021}, consisting of  a pointlike particle that decays into a pair of oppositely-propagating photons (Fig. \ref{decay}). 
  For a particle of mass $M$,  a spherical null gravitational shock wave from the photons creates a coherent anisotropic displacement of time at radius $R$, at an angle $\theta$ from the decay axis,   of magnitude 
 \begin{equation}\label{distortion}
 \delta \tau \simeq \frac{5GM}{24c^3} (3\cos^2\theta -1).
 \end{equation} 
  This distortion is the time displacement of clocks read in all directions by a single observer in the rest frame of the original decaying particle.
 The gravitational memory of the decay recalls  the whole history of a causal diamond: first the outgoing null shock, then the  incoming data from the clocks. 
 Since $\delta\tau$ is independent of $R$,  gravity  generates the same  coherent large-angle  pattern of distortions  on surfaces of causal diamonds  of any radius $R$.

 For  a  quantum decay where the axis is indeterminate, the correspondence principle demands that
 the space-time geometry is placed in a superposition of   different states with causal structures that share the same macroscopic angular distortion, apart from the axis direction.  
The clock displacements represent  a coherent measurement of the gravitational effect of a nonlocalized quantum state of matter--- the angular structure   of a quantum state of geometry measured ``from  inside''.   
The macroscopic coherent  distortion of  causal structure differs profoundly from na\"ive expectation of the QFT model that  states couple mainly on the de Broglie scale,
$\lambda \sim \hbar/Mc$. 


The same coherent effects generalize to a gas of gravitational shocks from a system with many null particles, and to vacuum fluctuations.
A  sum of many distortions like Eq. (\ref{distortion}) from more general quantum states of  null propagating pointlike particles leads to a coherent  large-angle  pattern with a similar angular power spectrum, but a larger amplitude.
Extrapolated to a  cosmological horizon volume,
 where the particles are numerous enough to create a mean curvature radius of order $R$, the large-angle gravitational distortions  $\Delta$ on causal surfaces of size $R$
 from a gas of  Planck mass particle particle states are of order\cite{Mackewicz2021}
  \begin{equation}\label{variance}
\langle \Delta^2\rangle = \langle \delta \tau^2\rangle/\tau^2 \simeq l_P /R,
\end{equation}
much larger than the value $\langle \Delta^2\rangle  \sim (l_P /R)^2$ predicted by QFT 
(but in agreement with a conformal field theory of near-horizon vacuum states with a Planck cut-off\cite{Banks2021}).
Such  large distortions  of causal surfaces from  vacuum-state fluctuations could even be detectable in direct experiments\cite{holoshear,Richardson2020}.
We are  led to suspect that  directionally coherent quantum gravity could dominate the production of perturbations, and possibly leave  distinctive universal signatures in the pattern  on large angular scales, where the curvature of cosmological horizons is important. 

Standard inflation theory
 beautifully  matches the angular power spectrum $C_\ell$ of CMB anisotropy\cite{WMAPfinal,Plancklegacy} on  angular  scales smaller than a few degrees, at angular wavenumbers  $ \ell\gtrsim 30$ where the difference between realizations, or  ``cosmic variance'', is relatively  small. Indeed, apart from the amplitude and tilt of the initial spectrum, the famous spectrum of acoustic peaks at  $\ell\gtrsim 100$  is shaped mostly by well-understood post-inflationary classical processes.  However, realizations of the large-scale pattern in the QFT model  vary  significantly,  depending  on random variations in initial conditions.  Thus, precise cosmological tests and parameter estimates often  omit data  on angular scales larger than a few degrees,    $\ell\lesssim 30$, which are the main focus here.

At large angles, the measured pattern  of the CMB temperature anisotropy, $\delta T/T$,  approximately preserves the primordial pattern  of gravitational distortions $\Delta$ on the spherical surface of last scattering\cite{Sachs1967}.
Since   conformal invariance of the expanding universe preserves  angular relationships, and physical horizons have spherical  boundaries  localized in comoving position,
   directional and causal constraints on inflationary horizons are more directly
related to  angular separation $\Theta$ than to angular wavenumbers $\ell$. 
Thus, we choose to study the pattern using the angular correlation function $C(\Theta)$  instead of its more familiar transform $C_\ell$. 
Defined  in terms of  angular averages,
 \begin{equation}\label{Ctheta}
C(\Theta) = \langle {\Delta}(\vec\Omega)   \langle {\Delta} \rangle_{\Theta,\vec\Omega}  \rangle_{\vec\Omega},
\end{equation}
where $ \langle \rangle_{\Theta,\vec\Omega}$  denotes an azimuthal mean on a circle at a  polar angle $\Theta$ about direction $\vec\Omega$, 
and $ \langle \rangle_{\vec\Omega}$ denotes a sky average.
 
\begin{figure*}[t]
\begin{centering}
\includegraphics[width=.49\linewidth]{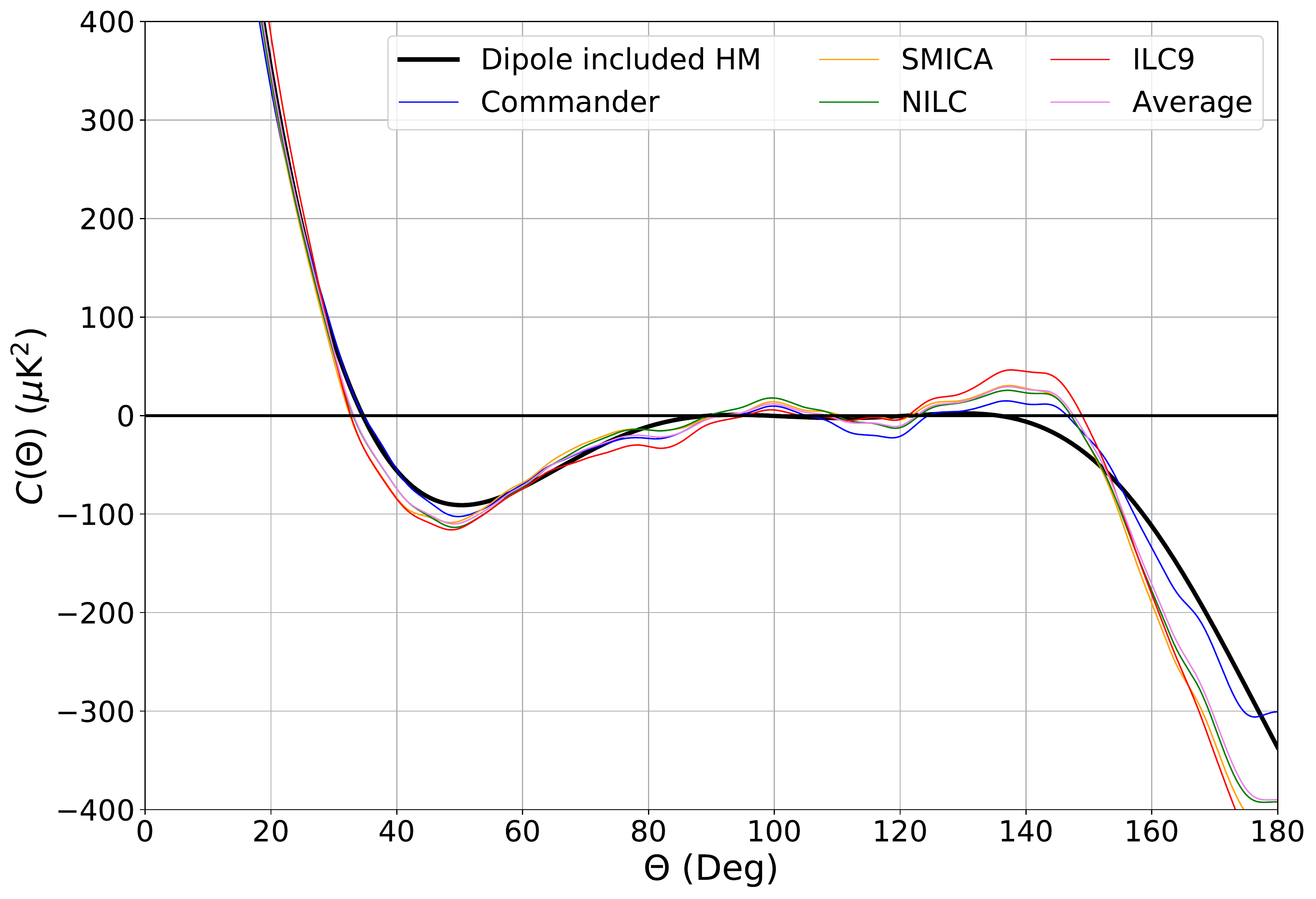}
\includegraphics[width=.49\linewidth]{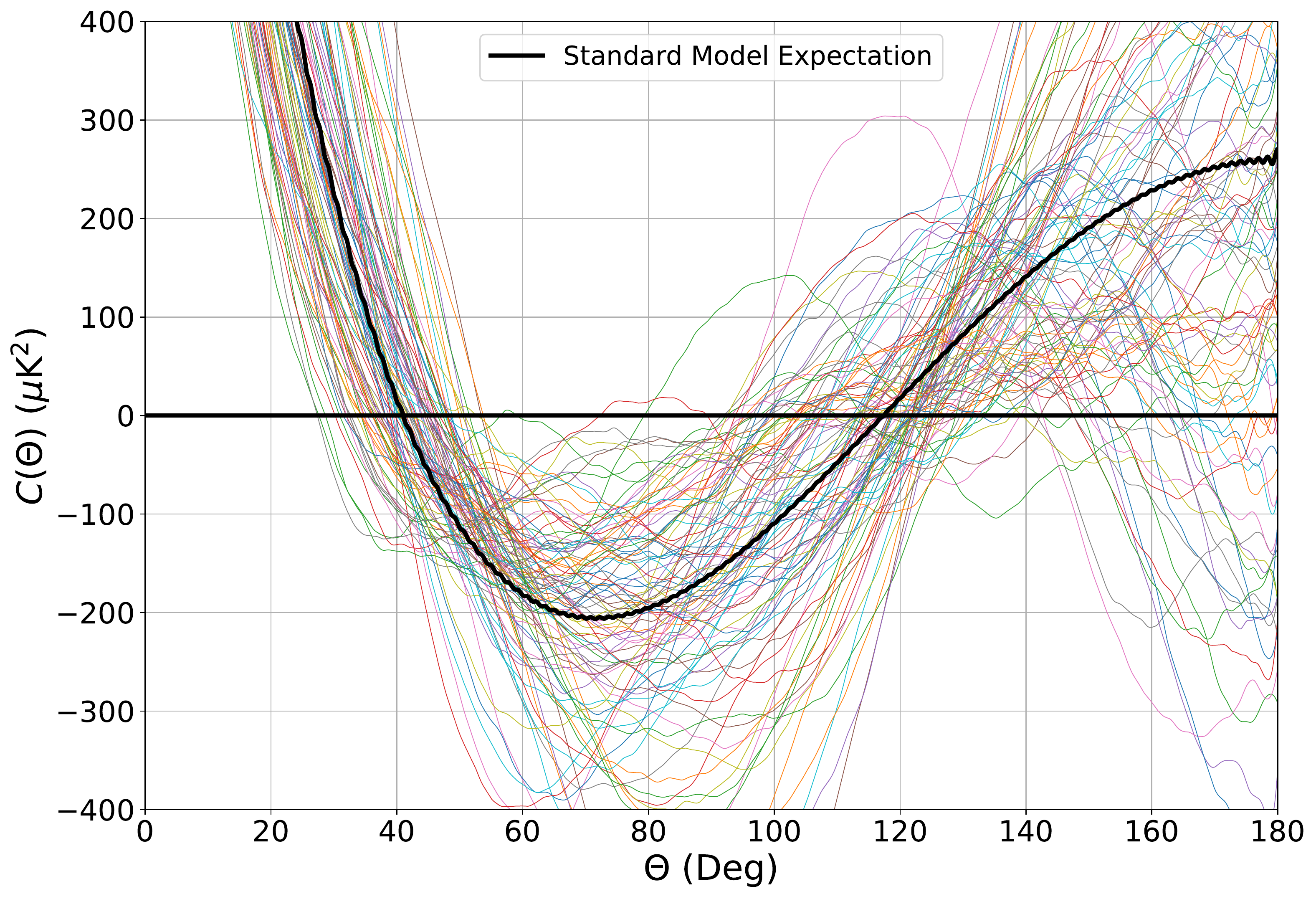}
\par\end{centering}
\protect\caption{Comparisons of measured CMB maps with models, reproduced from ref. \cite{Hogan_2022}. The colored lines in the left panel show the  correlation function $C(\Theta)$  of the real CMB sky (Eq. \ref{Ctheta}), as measured by the {\sl WMAP}  and  {\sl Planck} satellites\cite{WMAPfinal,planckforegroundsub2018}, after subtracting  various models of emission from our Galaxy.    The  ``holographic model'' (HM)  plotted\cite{Hogan_2022} is a simple analytic approximate model derived by integrating   dipolar distortions of directionally-entangled horizons, normalized to agree with the standard expected inflationary power spectrum on angular scales less than a few degrees.    All of these functions  are shown with  a $C_1\simeq 80\mu K^2 \cos(\Theta)$  contribution added to include the effect of  an intrinsic (but unobserved) dipole.   With this, the data appear to be consistent with zero for $90^\circ<\Theta<135^\circ$, as expected from causal symmetry; the spread among maps indicates the  current measurement uncertainty due to contamination by the Galaxy.
For comparison, on the right, 100  equally probable realizations in the standard QFT inflation model are shown on the same scale, none of which approximate the nearly-exact null symmetry that appears to describe the real sky. 
  The substantial large-angle variation is  mainly  from   wavelengths comparable to and  larger than the horizon, which may be an  artifact of  unphysically independent IR modes in the model.
  \label{data}}
\end{figure*}

It has been known for a long time that the measured  large-angle CMB angular correlation function  does  not appear to be at all typical of standard-model realizations\cite{WMAPanomalies,Ade:2015hxq,Akrami:2019bkn}.  
 In particular,  $|C(\Theta)|$  is much smaller than  expected from random cosmic variation at angular separations larger than about 80 degrees (Fig. \ref{data}).  Indeed, in the best  maps it appears to vanish near 90 degrees\cite{Hagimoto2020}, and 
is  consistent with zero over a range of larger angles, after allowing for measurement uncertainties and  for a small correction from an  intrinsic  unmeasured dipole\cite{Hogan_2022}.

\begin{figure*}
  \centering
  \includegraphics[width=0.9\textwidth]{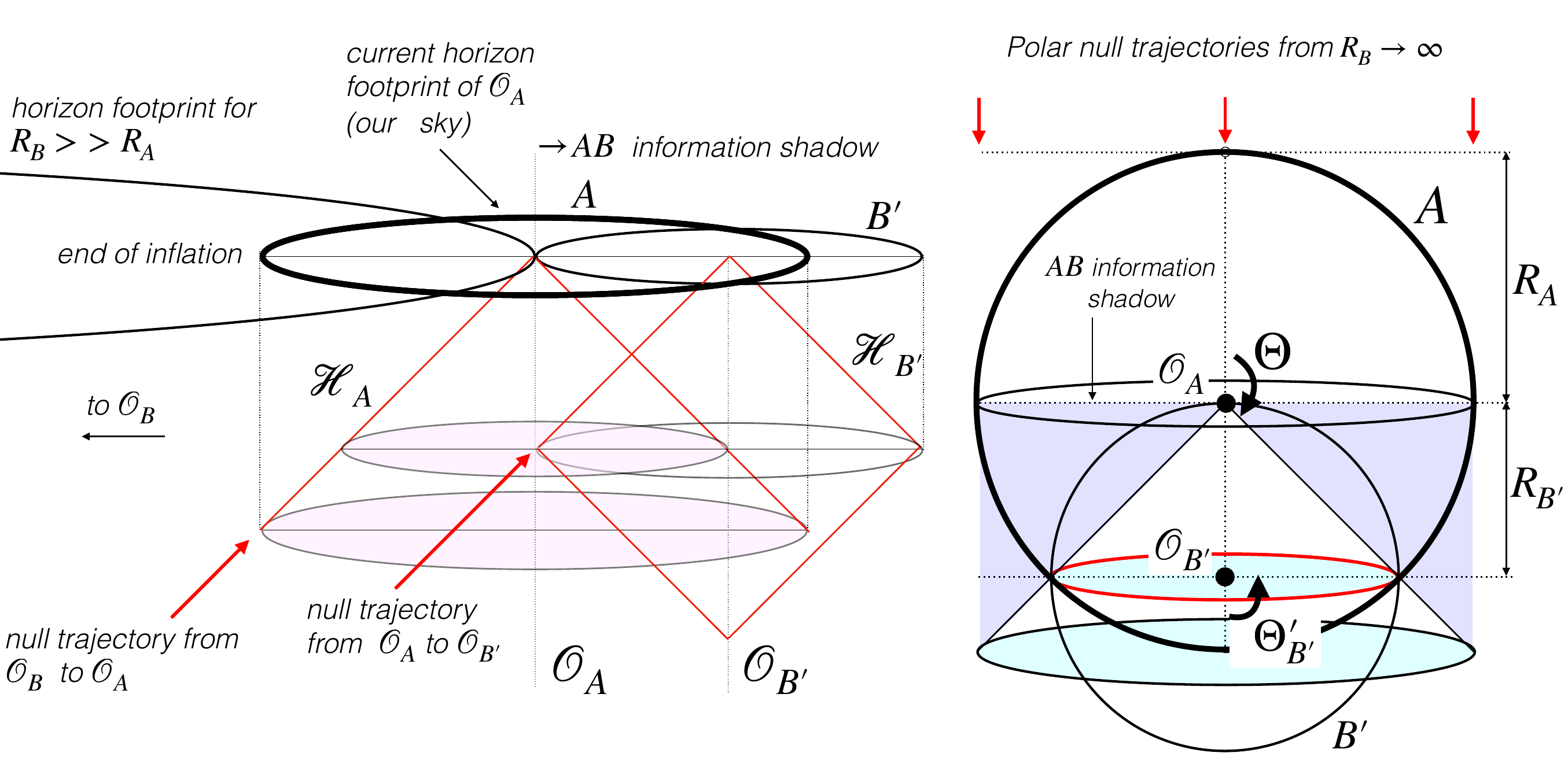}
  \caption{If cosmic perturbations  originate from directionally coherent quantum distortions of horizons,  boundaries of vanishing angular  correlation of gravitational potential among world lines are bounded by intersections of their causal diamonds during inflation.  The left panel shows a causal diagram of the inflationary era  in  comoving conformal coordinates\cite{Hogan_2022}: the vertical direction represents conformal time, and vertical lines represent comoving world lines.  Inflationary horizons ${\cal H}$  are  past null cones 
 that arrive at the end of inflation.  They intersect surfaces of constant conformal time on  ``horizon footprints'',  comoving 2D spherical surfaces of causal diamonds. Suppose  ${\cal O}_A$ lies on a horizon footprint  centered on ${\cal O}_B$, to the left in this figure.  Where $A$ and $B$  horizon footprints intersect, causal entanglement allows correlations.
 For $R_B\rightarrow \infty$, the $B$ footprint is a plane that intersects the $A$ equator: no information from any  point  on ${\cal O}_A$'s horizon reaches the antihemisphere on $A$  before inflation ends.  The  right panel shows  two horizon footprints in 3D comoving space at the end of inflation.  The polar angle $\Theta$ refers to angular correlations on our sky, the $A$  footprint.   The equator at $\Theta= 90^\circ$ defines the  boundary of the ``information shadow''  cast from the north polar direction.  The  $B'$ sphere shown, with $R_{B'}=R_A/\sqrt{2}$, bounds the   antihemispheric causal diamonds that can introduce nonzero correlation at $\Theta> 135^\circ$.\label{shadow}}
\end{figure*}


I will now argue that in a directionally coherent theory of quantum gravity,   such  a nearly-vanishing large-angle correlation has a natural interpretation:  it is a signature of 
{\it an exact null symmetry of the angular correlation function of  primordial perturbations},
\begin{equation}\label{exactnull}
C(90^\circ\le\Theta\le135^\circ) = 0;
\end{equation}
and that moreover,   this exact symmetry follows directly from
 constraints on causal relational information,  based on the hypothesis that  {\it primordial perturbations are  directionally   coherent quantum distortions of  causal diamonds}.   

Consider  null trajectories in comoving conformal coordinates during inflation (Fig. \ref{shadow}).  A direction in the sky is associated with a null trajectory on the inflationary horizon that arrives at our world line at the end of inflation.  Its 2D  normal plane represents the surface of a causal diamond of an infinitely distant point.   Correlated incoming  data from that direction intersects our sky  at a particular comoving location, a pole on the comoving 2D sphere (our ``horizon footprint'') approximately where the CMB last scattering surface is.  Information from the same infinitely distant point reaches the equator on the  same sphere only at the end of inflation, and never reaches the antihemisphere. {\it There is never a causal connection  between incoming directional data at a polar point  on our horizon and any point in its antihemisphere, so directionally coherent   correlations at $\Theta>90^\circ$  vanish. }

This ``information shadow'' argument relies on 
directional coherence  of  quantum gravitational distortions on causal diamonds,  similar to the directionally coherent gravitational memory in  the decaying-particle system  (Fig. \ref{decay}).  
It  does not apply  if the gravitational potential is quantized as a local scalar, as  in   QFT, where    
gravitational memory of a potential is ``frozen in'', mode by mode,  as a local scalar on world lines on surfaces of constant comoving time.
This reasoning highlights why incorrect  subtraction of the largest-scale modes in renormalized QFT matters:   directional  causal coherence leads to  nonlocal, in-common displacement of  $\Delta$ for whole causal diamonds relative to  infinitely distant points.

 The data also show  a  nonzero anticorrelation close to the antipode, that is,  a tendency for antipodal patches of sky to have opposite signs.
 This appears to be impossible according to the  classical causal reasoning just given, since no information from any direction on the horizon can directly reach its antihemisphere before the end of inflation.  However,  in a causally coherent theory of quantum gravity,  spacelike correlations can arise in the usual quantum-mechanical way, through quantum nonlocality within a coherent causal diamond. As in the original EPR thought-experiment, such correlations appear to be classically acausal, but actually show the causal effect of a nonlocal coherent state; as discussed above, the same coherence extends to  
 spacelike angular correlations of   gravitational time distortions.

 During inflation, nonzero  correlation  can arise near the antipode from  some  causal diamonds  smaller than  our horizon that  partially overlap, but extend beyond it (see Fig. \ref{shadow}).
As above, suppose there is a coherent in-common displacement of the potential of  a whole causal diamond in relation to any direction, that is, to an  infinitely distant point.
Points in the antihemisphere of a horizon footprint share this mean displacement, a kind of correlation that is invisible to the observer in the center.
 A causal diamond centered on a point ${\cal O}_{B'}$ in the antipodal direction, with ${\cal O}_A$ on its boundary,  also  inherits a coherent polar displacement from ${\cal O}_A$.   But, if ${\cal O}_A$ and  ${\cal O}_{B'}$ are neither too close nor too far apart,  anisotropy on the $B'$ horizon also creates correlation visible from ${\cal O}_A$.
The reciprocal $AB'$ causal relationship in the antihemisphere mirrors that used to derive the $AB$ information shadow;  thus, nonlocal spacelike correlations on $A$ can be introduced  if the  angular separation $\Theta'_{B'}=360^\circ-2 \Theta$ between the antipode  and the   $AB'$ horizon intersection circle, as viewed from ${\cal O}_{B'}$, lies in the range   $0<\Theta'_{B'}< 90^\circ$, or
 equivalently,   if information on a null plane  from ${\cal O}_A$ reaches ${\cal O}_{B'}$ before it reaches the $AB'$ horizon intersection.  As viewed from ${\cal O}_A$, these nonzero  correlations  appear for
$180^\circ>\Theta>135^\circ$,
which approximately agrees with  the real sky, after accounting for the unmeasured dipole.  
They reflect nonlocal correlations on $B'$  horizons over a limited range of sizes,
 $R_A/2<R_{B'}<R_A/\sqrt{2}$. 
Moreover, we expect that the
 parity-inverted relationship of $A$ and $B'$ horizons maps  positive small-angle correlation relative to $B'$  onto a negative large-angle correlation at the $A$ antipode,  
 \begin{equation}\label{anti}
C_{B'}(\Theta'_{B'}\rightarrow 0)>0\ \ \rightarrow \ \ C(\Theta\rightarrow 180^\circ)<0,
\end{equation}
again as observed.


An approximate analytic ``holographic'' model of $C(\Theta)$ constrained by such causal relationships\cite{Hogan_2022}, based on  integrating over  dipolar distortions of directionally-entangled  horizons with  relative amplitudes set by causal symmetries, is shown along with  data in Fig. (\ref{data}).  The overall amplitude and tilt of the 3D power spectrum are normalized to match the  angular power spectrum of standard cosmology where it works well, at separations smaller than a few degrees, but the holographic model matches the real  sky much better on large angular scales, with no additional adjustable parameters.

Model-independent comparisons\cite{Hogan_2022} show  that only a small fraction of standard realizations (between 1\% and 0.1\%, depending on the map)  approximate the real sky as well as the  large-angle null symmetry, Eq.( \ref{exactnull}).    In addition,  antipodal antisymmetry (Eq. \ref{anti})  corresponds to another  well-known measured anomaly of the large-scale angular spectrum,  a systematic excess of odd-parity over even-parity fluctuations for $\ell\lesssim 30$ that occurs in fewer than $1\%$ of realizations \cite{Akrami:2019bkn}.  
Other causal holographic symmetries can  probe information beyond that in the correlation function; for example,  previously-noted anomalies associated with alignments and shapes of low order multipoles may  signify a  universality of great circle variance, and reveal the orientation as well as the amplitude of the intrinsic dipole\cite{Selub2021}. 
Right now, the main obstacle  to  more powerful statistical tests of primordial angular symmetries appears to be  systematic uncertainty introduced by the foreground emission of our Galaxy, which can likely be improved.

According to the standard interpretation, these remarkable properties of the large-angle pattern are all statistical flukes.
As pointed out many times in the literature, a small value of $C(\Theta)$ at large angles in the standard model is not impossible,  because of the large variety of possible realizations;  however, {\it any exact null symmetry 
  of angular perturbations  is impossible}.
  The reason is plain to see in Fig. (\ref{data}):  independent, unconstrained  modes of QFT lead to random cosmic variation  inconsistent with the universal behavior required by a  directional symmetry.   To achieve an exact symmetry requires a nonlocal omnidirectional conspiracy among amplitudes and phases of modes on all scales, of the kind that would arise from the causality-enforced entanglement  described by Hollands and Wald.
Although any given outcome can be interpreted as a statistical fluke,  this particular outcome looks like a clue to something deeper going on, since  the measured shortcomings of  standard inflationary  predictions on large angular scales appear to   match   the theoretically known  shortcomings of the quantum field theory used to compute them.
Perhaps in the correct theory,  the apparent anomalies are  just  symmetries from causal constraints.



In a causally-coherent model, the  slow-roll behavior  of the unperturbed expected  inflationary background, as well as the successful match to the measured perturbation spectrum at $\ell>30$, which depends only on the slightly tilted power spectrum produced by the slow roll, remain the same as in the standard picture\footnote{Since the total variance now depends only on the  inflation rate $H$,    the effective potential that governs the classical slow-roll evolution must have a different form to match  constraints such as measured mean spectral tilt and required number of  inflationary $e$-foldings\cite{Hogan_2022}.}.
At the same time,   a fundamental  symmetry in $C(\Theta)$  at large angles provides concrete clues to the deep unsolved problem of how locality and coherence work in quantum gravity. If  symmetries of the sky  indeed represent physically fundamental constraints, the inflationary QFT approximation is unphysical.  Primordial gravitational perturbations  are not scalars attached to inflaton field mode fluctuations; instead,  they
 are coherent distortions of the emergent metric arising from quantum uncertainty in localization of causal surfaces.
 Coherent quantum states of geometry are not  localizable like quantum scalar particles;
  instead,  they live on null cones of emergent comoving world lines, so that
geometry retains quantum coherence on  horizons.

 A model of inflation with  coherent horizons resembles a holistic model of a black hole horizon, in which an entire 2+1D horizon surface is viewed as a single quantum object that creates nonlocal spacelike  entanglement with external states in all directions\cite{Hooft2016}.
 A cosmology based on this idea represents a radical conceptual departure from the standard view of initial conditions, since  its emergent definition of locality  allows no physical definition of a pre-existing, unperturbed classical background.  As exemplified by the decaying-particle thought-experiment,   the geometry ``becomes classical'' (that is, emerges from an indeterminate superposition into a  definite realization) only within each causal diamond.
In this view,  emergence of macroscopic space-time  from a quantum state continues even to the present day, as each world line's  comoving horizon  expands and  new information enters from outside.



\bibliography{GRFbib}

\end{document}